\documentclass[a4paper,11pt]{article}
\usepackage{pos}

\newcommand{\be}{\begin{equation}}
\newcommand{\ee}{\end{equation}}
\newcommand{\ba}{\begin{eqnarray}}
\newcommand{\ea}{\end{eqnarray}}

\newcommand{\rmO}{{\textrm{O}}}

\newcommand{\ahvp}{a_\mu^{\rm hvp}}

\newcommand{\bea}{\begin{eqnarray}}
\newcommand{\eea}{\end{eqnarray}}
\newcommand{\eq}[1]{eq.\,(\ref{#1})}

\title{Machine-learning techniques as noise reduction strategies in
  lattice calculations of the muon $g-2$}
\ShortTitle{Machine-learning techniques for $(g-2)_\mu$}

\author[a]{Thomas Blum}
\author[b,c]{Alessandro Conigli}
\author[d]{Lukas Geyer}
\author[e]{Simon Kuberski}
\author[b,c,d]{Alexander Segner}
\author*[d,b,c]{Hartmut Wittig}

\affiliation[a]{Department of Physics, University of Connecticut, Storrs, CT 06269, USA}

\affiliation[b]{Helmholtz-Institut Mainz, Johannes Gutenberg-Universit\"at Mainz, 55099 Mainz, Germany}

\affiliation[c]{GSI Helmholtz Centre for Heavy Ion Research, 64291 Darmstadt, Germany}

\affiliation[d]{PRISMA$^+$ Cluster of Excellence and Institut für
  Kernphysik, Johannes Gutenberg-Universit\"at Mainz, 55099 Mainz, Germany}

\affiliation[e]{Theoretical Physics Department, CERN, 1211 Geneva 23, Switzerland}

\emailAdd{hartmut.wittig@uni-mainz.de}
\abstract{Lattice calculations of the hadronic contributions to the
  muon anomalous magnetic moment are numerically highly demanding due
  to the necessity of reaching total errors at the sub-percent
  level. Noise-reduction techniques such as low-mode averaging have
  been applied successfully to determine the vector-vector correlator
  with high statistical precision in the long-distance regime, but
  display an unfavourable scaling in terms of numerical cost.  This is
  particularly true for the mixed contribution in which one of the two
  quark propagators is described in terms of low modes.  Here we
  report on an ongoing project aimed at investigating the potential of
  machine learning as a cost-effective tool to produce approximate
  estimates of the mixed contribution, which are then bias-corrected
  to produce an exact result. A second example concerns the
  determination of electromagnetic isospin-breaking corrections by
  combining the predictions from a trained model with a bias
  correction.
  \vspace*{0.5cm}
        \begin{flushright}
        CERN-TH-2025-038\\
        MITP-25-010
        \end{flushright}
}

\FullConference{The 41st International Symposium on Lattice Field Theory (LATTICE2024)\\
 28 July - 3 August 2024\\
Liverpool, UK\\}

\begin{document}
\maketitle

\section{Motivation}
\vspace{-0.2cm}

The hadronic vacuum polarization (HVP) contribution, $\ahvp$, to the
muon anomalous magnetic has been a major focus of lattice calculations
in recent years. The commonly used approach to compute $\ahvp$ is
based on the time-momentum representation \cite{Bernecker:2011gh}
which expresses $\ahvp$ as a convolution integral of the vector-vector
correlation function $G(t)$ multiplied by an analytically known kernel
function $\tilde{K}(t)$, i.e.
\begin{equation}\label{eq:TMRdef}
   \ahvp=\left(\frac{\alpha}{\pi}\right)^2\int_0^{\infty}dt\,\tilde{K}(t)G(t)\,.
\end{equation}
However, computing $\ahvp$ at the required level of precision is
numerically very costly due to the well-known ``noise problem'' which
manifests itself in the exponential growth of statistical errors in
$G(t)$ for large Euclidean times $t$
\cite{Parisi:1983ae,Lepage:1989hd}. Furthermore, in order to match or
exceed the precision of the traditional data-driven dispersive
approach, it is necessary to include isospin-breaking corrections
arising from QED and unequal up- and down-quark masses. While these
contributions are small, they are very expensive to calculate.

In this contribution, we explore the potential of machine-learning
methods to reduce the cost of evaluating computationally complex
and/or noisy correlation functions. The idea is to use a trained
neural net to produce approximate estimates for correlation functions
at low numerical cost, thereby predicting an ``expensive'' observable
through its correlation with a quantity that can be computed
``cheaply''. We report on first applications of this idea to compute
the vector two-point correlation function entering the TMR definition
of $\ahvp$, as well as the calculation of electromagnetic
isospin-breaking corrections to octet and decuplet baryon masses.

\section{Low-mode averaging and machine-learning strategy}
\vspace{-0.2cm}

\begin{figure}[t]
  \centering
  \includegraphics[width=0.8\textwidth]{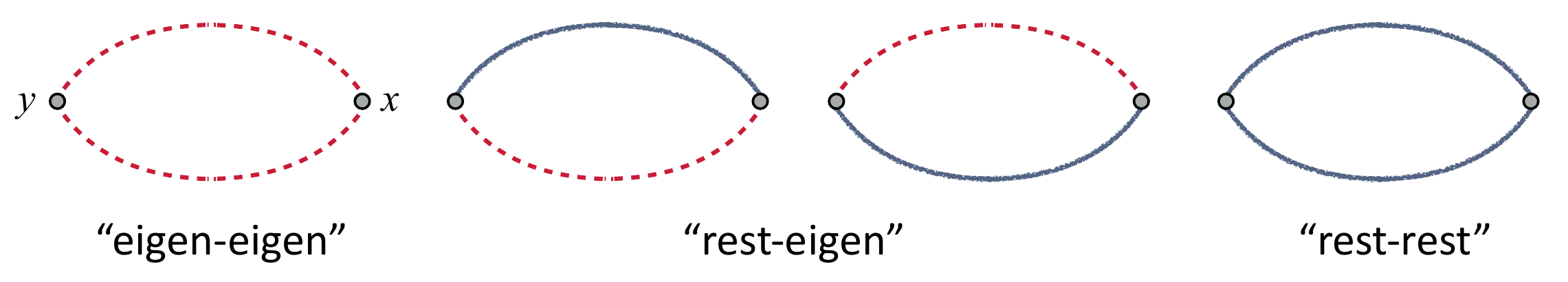}
  \caption{Sketch of the resulting two-point correlation function
    after applying the low-mode decomposition of
    \eq{eq:LMAdecomp}. Red dashed lines denote the low-mode part of
    the quark propagator expressed in terms of the spectral
    representation, while blue solid lines denote the high-mode
    part. The grey circles at points $x$ and $y$ are associated with a
    Dirac matrix.}
  \label{fig:LMAdecomp}
\end{figure}

Deflation techniques such as low-mode averaging (LMA)
\cite{Giusti:2004yp,DeGrand:2004qw} have been applied successfully to
mitigate the noise problem, e.g. in calculations of the HVP
contribution using Wilson fermions~\cite{Djukanovic:2024cmq}. LMA is
based on splitting the quark propagator into a low-mode contribution
and a part defined in the orthogonal complement of the space spanned
by the chosen number of low eigenmodes of the system, according to
\be
  S(y,x) = S_{\rm eigen}(y,x)+S_{\rm rest}(y,x)\,,
\ee
where the subscripts ``eigen'' and ``rest'' denote the low- and
high-mode parts, respectively. If we denote the chosen number of
eigenmodes by $N_{\rm low}$, then the low-mode part of the propagator
can be written in terms of the spectral decomposition as
\be\label{eq:LMAdecomp}
  S_{\rm eigen}(y,x)=\sum_{i=1}^{N_{\rm low}}
  \frac{v_i(y)\otimes(\gamma_5 v_i(x))^\dagger}{\lambda_i},\quad
  (\gamma_5 D_{\rm w})\,v_i(x)=\lambda_i v_i(x)\,,
\ee
where $v_i(x)$ is the eigenmode of the hermitian Wilson-Dirac operator
$\gamma_5 D_{\rm w}$ with eigenvalue $\lambda_i$. When computing
two-point correlation functions using the decomposition of
\eq{eq:LMAdecomp}, one obtains the ``eigen-eigen'', ``rest-rest'' and
``rest-eigen'' contributions schematically shown in
Fig.\,\ref{fig:LMAdecomp}. In the case of the vector-vector correlator
$G(t)$ which is relevant for determining the HVP contribution, one
finds that the ``eigen-eigen'' part dominates the long-distance
regime, as expected (see Appendix~C of Ref.\,\cite{Djukanovic:2024cmq}
for a detailed account). While the mixed rest-eigen contribution is
sub-dominant in the long-distance regime, its contribution to the
variance can be sizable. Its evaluation scales with the number of
eigenmodes, which substantially contributes to the cost, since
typically $N_{\rm low}=\rmO(1000)$. Hence, if the rest-eigen part were
accurately predicted through a machine-learning approach, it would
potentially reduce the computational cost by a big margin.

Our machine-learning approach is inspired by the well-known
all-mode-averaging (AMA) technique \cite{Blum:2012uh}. The goal is to
compute many approximate estimates, ${O}_{\rm appx}$, at low numerical
cost and obtain an exact result for the quantity of interest after
applying a bias correction, i.e.
\be
  \left\langle O\right\rangle =
  \left\langle{O}_{\rm appx}\right\rangle+
  \left\langle(O-O_{\rm appx})\right\rangle\,.
\ee
While the calculation of the correction $(O-O_{\rm appx})$ involves
the exact (and hence more expensive) evaluation of the observable $O$,
it is performed much less frequently than the computation of ${O}_{\rm
  appx}$. Provided that the approximate estimate can be computed with
a smaller variance than the exact evaluation and that the error of the
bias correction is sub-dominant, the procedure will be numerically
more efficient. In the familiar case of AMA, the approximate solution
is provided by the truncated solver technique; here we will replace
it by a trained model.
The idea to use trained networks for producing approximate estimates
for an observable followed by a bias correction has been tried in the
context of baryonic two- and three-point functions several years ago
\cite{Yoon:2018krb}. Here we will explore the potential of this method
for obtaining statistically accurate results for the vector correlator
$G(t)$ at large times. Specifically, we will train a model to predict
the numerically expensive sub-leading rest-eigen contribution, given
the eigen-eigen and rest-rest contributions as input.

\begin{table}
  \centering
  \begin{tabular}{cccccc}
    \hline\hline
    Id & $L/a$ & $T/a$ & $a\,[\rm fm]$ & $m_\pi$ & $N_{\rm cfg}$ \\
    \hline
    A654 & 24 & ~48 & 0.097 & 338 & 2500 \\
    D450 & 64 & 128 & 0.075 & 219 & ~500 \\
    N451 & 48 & ~96 & 0.075 & 291 & 1011 \\
    \hline\hline
  \end{tabular}
  \caption{CLS ensembles used in our project. Ensembles A654 and D450
    were used to predict the rest-eigen contribution to pseudoscalar
    and vector two-point functions. Electromagnetic corrections to
    baryon masses were computed on ensemble N451.}
  \label{tab:ens}
\end{table}

\section{Pseudoscalar and vector two-point correlation functions}
\vspace{-0.2cm}

Our pilot study was performed on two CLS ensembles, A654 and D450, of
O($a$)-improved Wilson fermions, both with periodic boundary
conditions in time. Some of their features are listed in
Table~\ref{tab:ens} together with those of a third ensemble (N451)
used in our second test on predicting electromagnetic corrections to
baryon masses. For each ensemble we divided the set of configurations
into subsets for training, testing and bias correction. While the sets
for testing and bias correction may share configurations, the training
set is completely disjoint.

We conducted extensive tests on multiple common approaches for
regression tasks, spanning from simple linear regression models to
deep neural networks. This resulted in a large set of models with
increasing complexity, where the number of trainable parameters varied
from $\rmO(100)$ to $\rmO(10^5)$. We systematically explored the
hyper-parameter space of our models using a grid-search algorithm,
assessing the performance of each model through cross-validation on
the training set. Our analysis suggests that architectures with a
larger number of parameters might suffer from overfitting,
particularly when the number of training epochs is increased. On the
other hand, simple linear regression models typically struggle to
capture the underlying correlation in our data, resulting in poorer
predictions. In addition, we explored two different approaches: one
where a single model predicts all timeslices simultaneously and
another where separate models are trained for each individual time
slice. Our analysis revealed comparable performance between these two
strategies. For the rest of this study, we adopt the full timeslice
prediction approach. Considering the results of our grid-search and
the relatively small size of our training set compared with typical
machine-learning applications, we employ a fully connected network
with one hidden layer. The hidden layer employs ReLU activation
functions to overcome the vanishing gradient problem, while the output
layer uses a linear activation function to maintain unbounded
predictions. To mitigate overfitting, we include dropout layers
\cite{hinton2012improving}, which randomly deactivate a subset of
parameters during training, thereby enhancing model generalization.
To evaluate the performance of regression models, we used the mean
squared error as our loss function. We found that normalizing the
training set to have mean zero and standard deviation can
significantly improve the network performance, leading to faster
convergence \cite{LeCun2012}. Moreover, we explored variations of the
training set by incorporating either the eigen-eigen data, the
rest-rest data, or a combination of both into our models to evaluate
their impact on performance. We find that a larger training set
containing both the rest-rest and eigen-eigen data leads to improved
network performance.

\begin{figure}[t]
  \centering
  \includegraphics[width=0.97\textwidth]{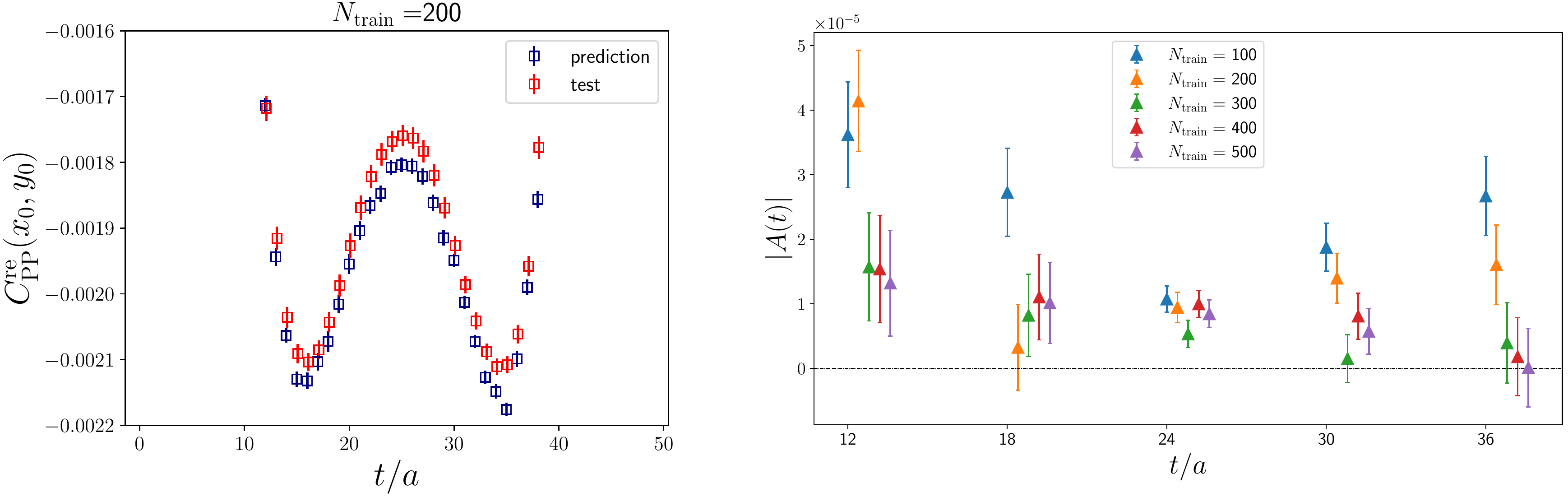}
  \vspace{-0.3cm}
  \caption{{\em Left:} The rest-eigen part of the pseudoscalar
    correlator on ensemble A654 as predicted by the model compared
    with the exact calculation for $N_{\rm train}=200$. {\em Right:}
    Deviation between prediction and exact calculation for different
    sizes of the training set. \label{fig:PS-A654-train}}
  \vspace{0.5cm}
  \includegraphics[width=0.97\textwidth]{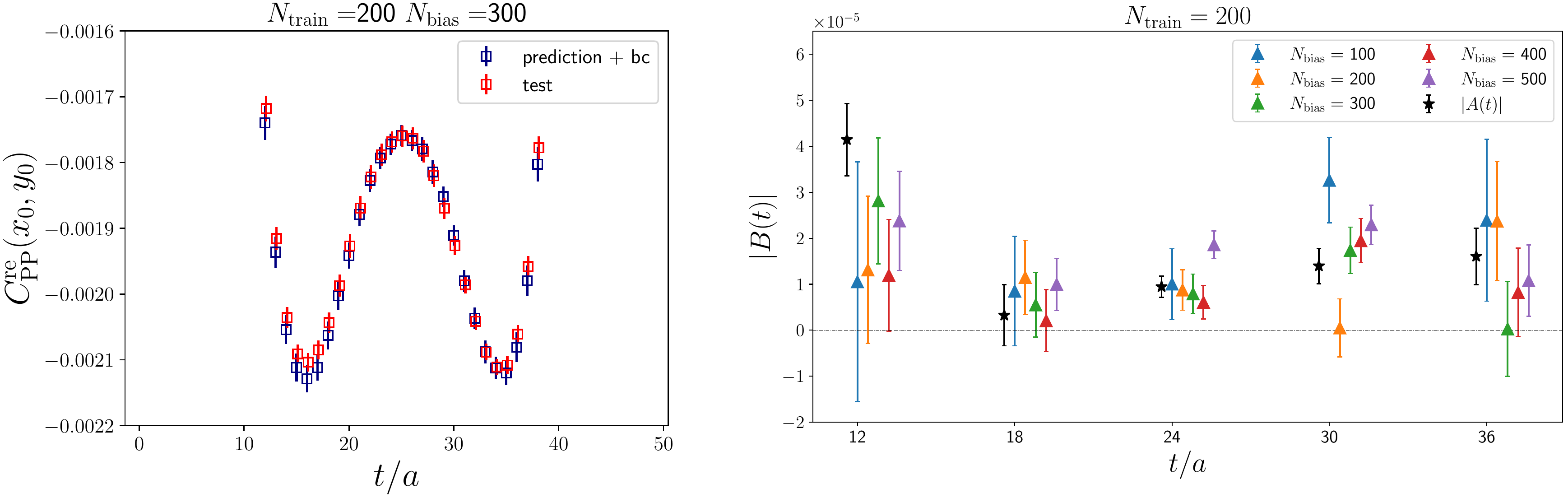}
  \vspace{-0.3cm}
  \caption{{\em Left:} The bias-corrected rest-eigen part of the
    pseudoscalar correlator on ensemble A654 with the exact
    calculation for $N_{\rm train}=200,\,N_{\rm bias}=300$. {\em
      Right:} Bias correction $B(t)$ for different values of $N_{\rm
      bias}$. The deviation $A(t)$ between prediction and exact
    evaluation is shown as the black star. \label{fig:PS-A654-bias}}
\end{figure}

Before studying the more relevant case of the vector correlator, we
applied our model to predict the rest-eigen contribution of the
pseudoscalar correlation function which does not suffer from the noise
problem. In Fig.\,\ref{fig:PS-A654-train} we show the rest-eigen
contribution in the central region. The left panel compares the
prediction from the model with the exact evaluation of the correlator
on an identical set of configurations. For this comparison, 200
configuration were used in the training step. The right panel shows
the quality of the prediction as a function of the size of the
training set, $N_{\rm train}$. The quality is expressed in terms of
the absolute value of the quantity
\be
  A(t)=\left\langle C^{\rm exact}(t)\right\rangle_{\rm test}-
  \left\langle C^{\rm pred}(t)\right\rangle_{\rm test}\,,
\ee
which is a measure of the deviation between model prediction and exact
calculation at timeslice $t$, both evaluated on the entire test set of
configurations. While there is a tendency that larger training sets
yield a better prediction, the values of $A(t)$ are quite stable when
$N_{\rm train}\gtrsim200$. The left panel of
Fig.\,\ref{fig:PS-A654-bias} shows that the bias-corrected prediction
agrees very well with the exact calculation. In the right panel the
absolute value of the bias correction, defined by
\be
   B(t)=\left\langle C^{\rm exact}(t)-
        C^{\rm pred}(t)\right\rangle_{\rm bias}\,,
\ee
is shown for different values of $N_{\rm bias}$ and compared against
the mismatch $A(t)$ between prediction and exact calculation. The fact
that the magnitudes of $A(t)$ and $B(t)$ agree well within errors
further illustrates the ability of our machine-learning model to
produce accurate results for the rest-eigen part of the pseudoscalar
correlator.

\begin{figure}[t]
  \centering
  \includegraphics[width=0.97\textwidth]{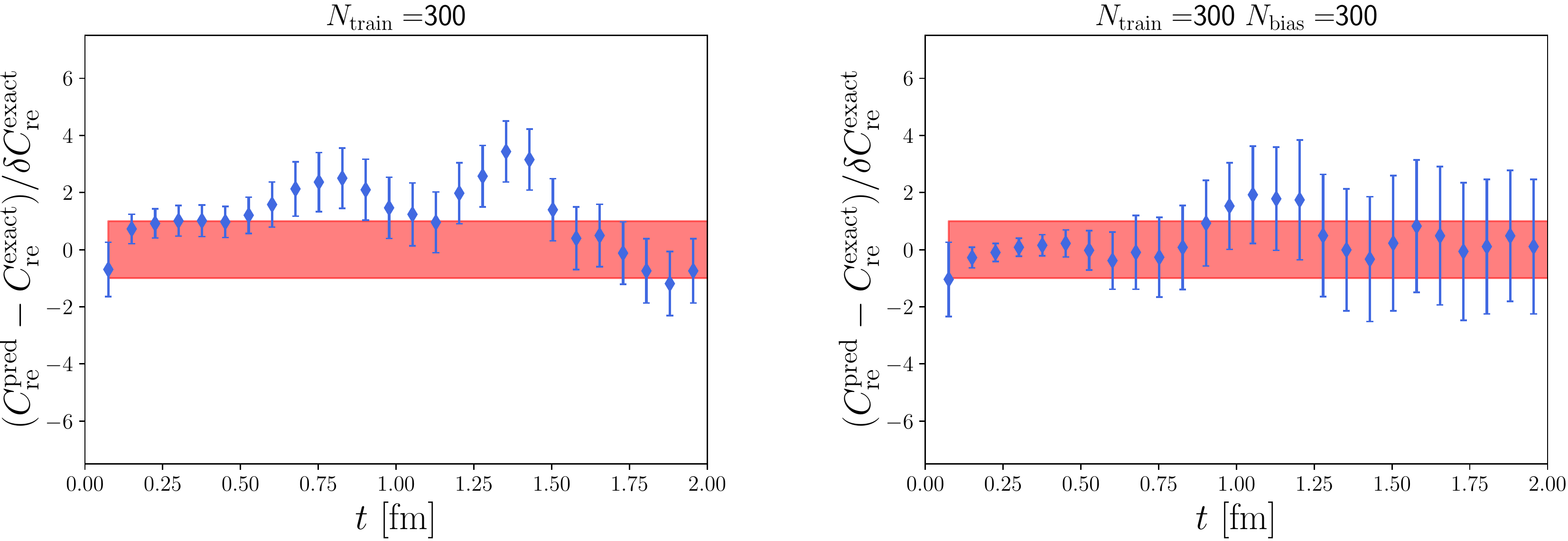}
  \vspace{-0.3cm}
  \caption{{\em Left:} Deviation between the predicted and exact
    rest-eigen part of the vector-vector correlator $G(t)$ in units of
    the statistical error of the exact result, plotted versus
    Euclidean time. The red horizontal band denotes the one-sigma
    error. {\em Right:} Bias-corrected prediction for the
    vector-vector correlator. \label{fig:VV-A654}}
  \vspace{0.5cm}
  \includegraphics[width=0.97\textwidth]{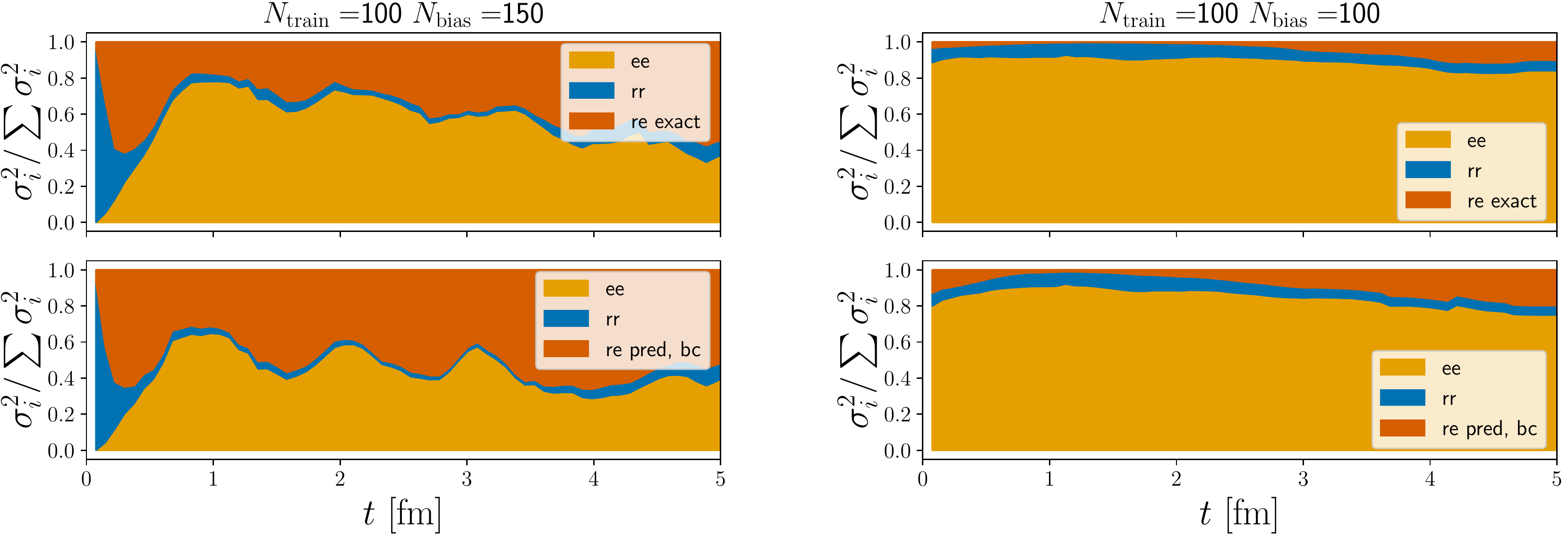}
  \vspace{-0.3cm}
  \caption{Fractional contributions of the eigen-eigen (ee),
    rest-eigen (re) and rest-rest (rr) parts to the total error of the
    correlator as a function of Euclidean time for ensemble D450. The
    upper plot in each panel shows the exact calculation with the
    machine-learning model shown in the lower plots. The vector
    correlator is shown in the two panels on the left, while the
    pseudoscalar case is shown on the right. \label{fig:VV-PS-D450}}
\end{figure}

Moving to the case of the (light-quark connected) vector-vector
correlator, we plot in Fig.\,\ref{fig:VV-A654} the deviation between
prediction and exact calculation of the rest-eigen part, with and
without bias correction, in units of the statistical error of the
exact result. While the quality of the prediction for the vector
correlator is similar to that of the pseudoscalar one (left panel),
one finds that the statistical error of the bias-corrected result is
about twice as large as for the exact calculation at large Euclidean
times (right panel). This much more unfavourable behaviour compared to
the pseudoscalar case is also illustrated by
Fig.\,\ref{fig:VV-PS-D450} where we compare the fraction of the
variance contributed by the eigen-eigen, rest-eigen and rest-rest to
the total uncertainty of the correlator. The right panel of the figure
shows that the eigen-eigen contribution dominates the error in the
long-distance regime of the pseudoscalar correlator, whereas the
largest contribution to the uncertainty of the vector correlator comes
from the rest-eigen part. A reduction of the total error in the
bias-corrected prediction of the vector correlator is possible but at
the expense of increasing the number of configurations used in the
bias correction to a level where the numerical gain
%of the machine-learning approach
is all but gone.

\section{Isospin-breaking corrections to baryon masses}
\vspace{-0.2cm}

\begin{figure}[t]
  \centering
  \includegraphics[width=0.97\textwidth]{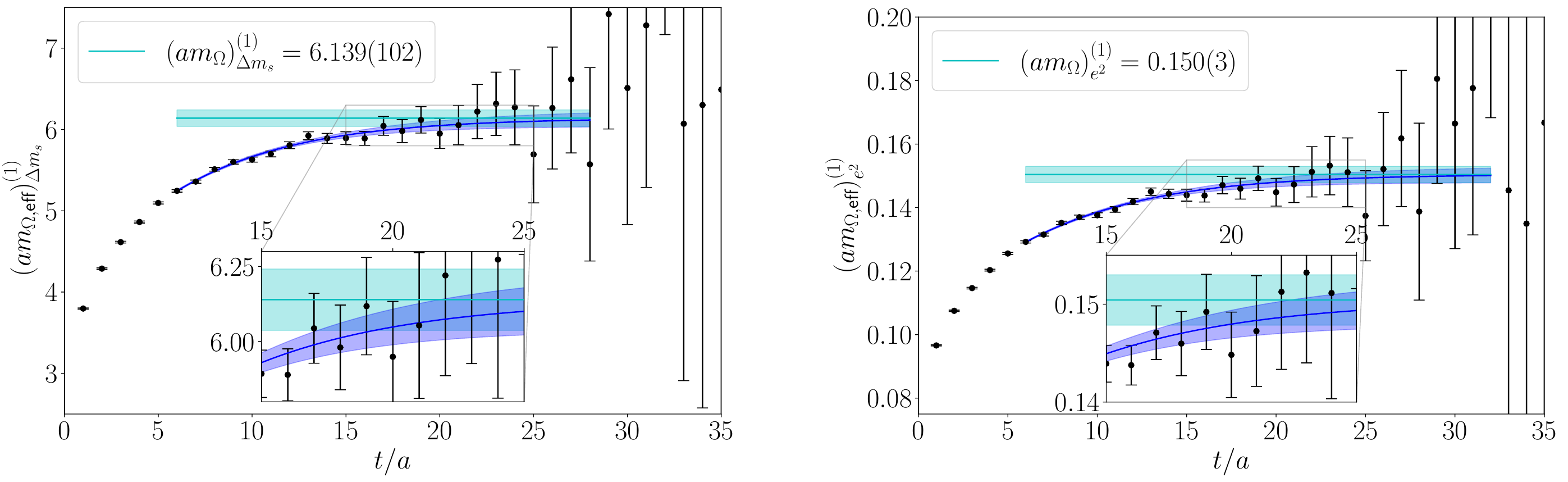}
  \vspace{-0.3cm}
  \caption{The contributions from the strange quark mass detuning
    (left panel) and electromagnetic corrections (right panel) to the
    effective mass of the $\Omega^-$ baryon, in lattice
    units. \label{fig:SIB-QED}}
  \vspace{0.5cm}
  \includegraphics[width=0.97\textwidth]{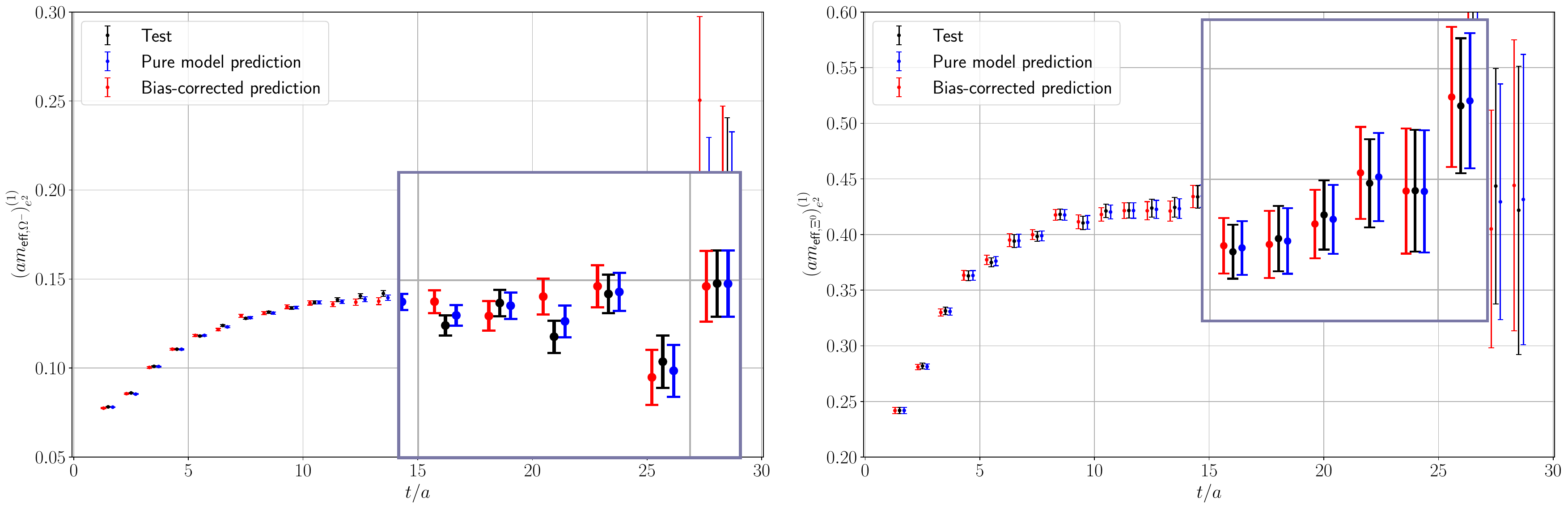}
  \vspace{-0.3cm}
  \caption{Electromagnetic corrections to the effective mass of the
    $\Omega^-$ (left panel) and $\Xi^-$ (right panel) baryons plotted
    in lattice units. The insets show the enlarged region for
    $t/a=16-21$. \label{fig:baryons}}
\end{figure}

Lattice calculations of the HVP contribution to the muon $g-2$ with
sub-percent precision require that the uncertainty in the lattice
scale be about two times smaller than the target precision, in order
for the scale setting error to be
sub-dominant~\cite{DellaMorte:2017dyu}. The masses of the lowest-lying
octet and decuplet baryons have emerged as reliable quantities for
scale setting, since these states are stable in QCD and can be
computed with small statistical errors. For the long-term goal of
determining the Standard Model prediction to the muon $g-2$ with a
precision that can rival the experimental measurement, it is necessary
to include electromagnetic corrections as well as corrections due to
the mass splitting between up and down quarks. In
Ref. \cite{Segner:2023igh} we have presented first results from an
ongoing project to compute octet and decuplet baryons including strong
and electromagnetic isospin-breaking effects. To this end we employ
the approach pioneered by the RM123 collaboration
\cite{deDivitiis:2011eh,deDivitiis:2013xla} which relies on an
expansion about iso-symmetric QCD. Full details about the
implementation of this formalism in the context of baryonic two-point
correlation functions can be found in \cite{Segner:2023igh}. The
numerical evaluation of baryon masses in QCD+QED via this approach is
numerically very demanding, with 50\% of the computer time spent on
computing the relevant correlators including photon lines. However, as
can be seen from Fig.\,\ref{fig:SIB-QED}, one finds that
electromagnetic isospin-breaking corrections are strongly correlated
with the quark mass detuning (i.e. the difference between the quark
masses defining QCD+QED and iso-symmetric QCD, respectively). This
suggests a machine-learning strategy that is based on training a model
on the correlation between strong and electromagnetic isospin-breaking
corrections to predict the latter and apply a bias correction to
produce an exact result. We consider a simple linear model, i.e.
\be\label{eq:QEDmodel}
   M(t)=\alpha(t)\,C^{(0)}(t)+\beta(t)\,C^{(1)}_{\Delta m_u}(t)
+\gamma(t)\,C^{(1)}_{\Delta m_d}(t)
+\delta(t)\,C^{(1)}_{\Delta m_s}(t)
+\epsilon(t)\,,
\ee
where $C^{(0)}(t)$ denotes the baryon correlation function in
iso-symmetric QCD, while $C^{(1)}_{\Delta m_q}(t),\,q=u,\,d,\,s$
denote the contributions arising from the quark mass detuning. The
parameters $\alpha,\,\beta,\ldots,\,\epsilon$ are determined in the
training step such that the model predicts the QED contribution
$C^{(1)}_{e^2}(t)$ given $C^{(1)}_{\Delta m_q}(t)$ as input.
To this end we divided ensemble N451 into a training set of $N_{\rm
  train}=20$ configurations and a test set of size $N_{\rm test}=991$.
For each configuration from the training set we performed the exact
calculation of the QED contribution for all of the 32 quark sources
used in our setup. For each timeslice~$t$ we then applied the linear
regression of the QED contribution according to \eq{eq:QEDmodel}. The
saved parameters $\alpha(t),\ldots,\epsilon(t)$ from the training step
were then used to evaluate the QED contribution via \eq{eq:QEDmodel}
for each configuration of the test set and for all of the 32 quark
sources. For the bias correction, we evaluated the exact QED
contribution on all configurations of the test set, but only for a
single quark source, thereby reducing the computational effort for
evaluating electromagnetic corrections by a factor~32. Given that QED
corrections account for about 50\% of the total time, this produces a
numerical gain by a factor~2. We found that the combined statistical
errors of the prediction and the bias correction are competitive with
the exact calculation for all 32 quark sources. This is seen clearly
in Fig.\,\ref{fig:baryons} which shows the electromagnetic corrections
to the $\Omega^-$ and $\Xi^-$ effective masses. For both states, the
bias-corrected predictions agree with the exact calculation within
errors which are also of comparable size. We have studied the effect
of increasing the number of sources in the bias correction and found
that the errors stayed approximately constant.

\section{Conclusions}
\vspace{-0.2cm}
Machine-learning techniques offer a viable alternative to conventional
methods for calculating correlation functions in lattice QCD, although
their efficiency and practicality depend on the specific
application. In our study of isospin-breaking corrections to baryon
masses, we observed a strong correlation between the electromagnetic
corrections and quark mass detuning. These correlations enabled us to
construct a model that achieved substantial resource savings in
determining electromagnetic corrections, reducing the computer time
required to reach our targeted precision by approximately 50\%.
In contrast, the analysis of the rest-eigen contribution to the vector
correlator $G(t)$ in the long-distance regime is less
straightforward. Although our machine-learned correlator, combined
with a bias correction, can reproduce $G(t)$, the uncertainty
remains larger than that obtained from an exact evaluation, with the
bias correction step being the dominant source of error and
computational expense. This outcome suggests that our model has not
yet fully exploited the correlations among the various contributions
to $G(t)$, highlighting an important area for future improvement.
We will continue the search for alternative models designed to
increase the correlations between the approximate and exact evaluation
of the rest-eigen contribution.
If successful, our search may lead to substantial savings. Typically,
the fraction of CPU time required to compute the rest-eigen
contribution is 50\%. Since the numerical overhead for the training
effort is negligible, one can hope for a 50\% reduction in CPU time in
the best of all cases.

\section*{Acknowledgements}
\vspace{-0.2cm}
Calculations for this project have been performed on HPC platforms at
Helmholtz Institute Mainz, Johannes Gutenberg-Universit\"at Mainz,
J\"ulich Supercomputing Centre (JSC) and
H\"ochstleistungsrechenzentrum Stuttgart (HLRS).  We gratefully
acknowledge the support of the Gauss Centre for Supercomputing (GCS)
and the John von Neumann-Institut f\"ur Computing (NIC) for projects
HMZ21 and HINTSPEC at JSC and project GCS-HQCD at HLRS.
This work has been supported by the German Research Foundation through
the Cluster of Excellence ``Precision Physics, Fundamental
Interactions and Structure of Matter'' (PRISMA+ EXC 2118/1), funded
within the German Excellence strategy (Project ID 39083149). SK is
supported by the European Union's Horizon Europe research and
innovation programme under the Marie Sk\l{}odowska-Curie grant
agreement No.\ 101106243. TB is partially supported by the US DOE
under grant DE-SC0010339.

%\bibliographystyle{JHEP}
%\bibliography{biblio}

\providecommand{\href}[2]{#2}\begingroup\raggedright\endgroup

\end{document}